# Generation of arbitrary radially polarized array beams by modulating the correlation structure

Shijun Zhu,[a] Jing Wang, and Zhenhua Li

*Department of Information Physics and Engineering, Nanjing University of Science and Technology, Nanjing, 210094,*

*China*

We demonstrate a convenient approach for simultaneously manipulating the amplitude and polarization of light beams by means of the modulation of the correlation structure. As an illustration, we constructed a periodic correlation structure that can generate an arbitrary radially polarized array (RPA) beam of a radial or rectangular symmetry array in the focal plane from a radially polarized (RP) beam. The physical realizability conditions for such source and the far-field beam condition are derived. It is illustrated that the beamlet shape and the state of polarization (SOP) can be effectively controlled by the initial correlation structure and the coherence width. Furthermore, by designing the source correlation structure, a tunable OK-shaped RPA beam and an optical cage are demonstrated, which can find widespread applications in non-destructive manipulation of particles and living biological cells. The arbitrariness in the design of correlation structure prompted us to find more convenient approaches for controlling the statistics of light beams in terms of amplitude and polarization. @ *2016 AIP Publishing LLC*.

In the past decade, array beams have seen a rapid growth of interest in the use of optical interconnections, ultracold atoms and optical tweezers [1–4]. Therefore, a variety of approaches have been introduced for generating array beams such as arrays of microlenses, binary phase gratings, and interferometric technique [5–7]. In general, array beams in previous studies including Gaussian array beams, dark-hollow array beams, flat-topped array beams, and Bessel array beams with a spatially homogeneous state of polarization (SOP), are within the framework of scalar beams [8–10]. However, if the SOP of a light beam is inhomogeneous but is instead a spatially varying polarization distribution in the transverse beam cross-section, vector theory is needed. Due to their essential features that are not present in scalar beams, vector beams have started to open up new research opportunities in optics [11].

---

[a] Author to whom correspondence should be addressed. Electronic mail: shijunzhu@njust.edu.cn.

On the other hand, significant efforts have been devoted to the generation of cylindrical vector (CV) beams, a family of vector solutions of Maxwell's equation, which features a cylindrical SOP orthogonal to the propagation direction. As an important class of CV beams, radially polarized (RP) beams have been extensively investigated due to their special features such as a tight focal beam spot and strong longitudinal component of the electric field at the focal plane, which have proved to be a powerful tool for many applications in polarization information encryption, optical data storage, super-resolution imaging, material processing, plasmonic focusing and optical tweezers [12–23]. Various methods for generating RP beams in terms of interferometric technique, modified laser cavities, binary optical elements, half-wave plates, and liquid crystal phase modulators have been widely explored [24–32]. However, very few work has referred to the generation of radially polarized array (RPA) beams. In some cases of practical application, for instance, multiple manipulation of nanoparticles, living biological cells and organelles within cells, a RPA beam have significant advantages in comparison with conventional scalar array beams [3, 12].

The notion of coherence is an important fundamental of modern optics. It is well known that the coherence is intrinsically connected with the spectral intensity, SOP, beam quality and the orbital angular moment (OAM) [33–42]. For instance, one cannot determine coherence from the visibility of interference pattern when the SOP of a vector beam is ignored. Recently, a sufficient condition was introduced by Gori *et al*. for devising genuine spatial correlation functions based on the non-negative definiteness constraint of the cross-spectral density (CSD) [43, 44]. A variety of light beams with peculiar correlation properties have been proposed, for which unexpected features have been revealed such as self-accelerating, ring-shaped profile as well as optical lattices in the far field [45–56]. In this paper, we extend the work from scalar case to the vector case, and demonstrate the ability of correlation structure to simultaneously manipulate the amplitude and polarization. By constructing a tunable periodic correlation structure, we have achieved a physically feasible approach for converting a RP beam into an arbitrary RPA of highly directional beam in the focal plane. The realizability conditions for such source and the beam condition for radiation generated by such source are established. It has been shown that the far-field SOP and the beam shape can be effectively manipulated by taking advantage of the source correlation structure and the coherence width. As further explorations, a tunable OK-shaped RPA beam and an optical cage are demonstrated by using a similar approach, which will be found important applications in non-destructive manipulation of living biological cells and capture of metal nanoparticles. The novelty of manipulating the amplitude and polarization lies in replacing the complex phased array design by a convenient coherence structure modulation based on the unified theory of coherence and polarization.

In general, the spatial coherence properties of a light field can be characterized by means of CSD. Within the paraxial approximation yield, the second-order correlation properties of a quasi-monochromatic, statistically stationary electromagnetic beam propagating along z-direction can be described by the $2\times 2$ CSD matrix in space-frequency domain, whose elements are given by [33]

$$\Gamma_{\alpha\beta}(\mathbf{r}_1,\mathbf{r}_2)=\langle E_\alpha^*(\mathbf{r}_1)E_\beta(\mathbf{r}_2)\rangle; \quad (\alpha,\beta=x,y), \tag{1}$$

where $\mathbf{r}_j \equiv (x_j',y_j')$, $(j=1,2)$ is the arbitrary position vector in the source under the Cartesian coordinate. The angle brackets denote time average and the asterisk represents the complex conjugate. $E_\alpha(\alpha=x,y)$ denotes the fluctuating electric field component along α-direction. For simplicity, we not specify the frequency $\omega$ throughout this paper.

It is well known that the coherence is closely associated with the polarization. Naturally, one can expect to modulate the polarization by means of the coherence. However, although the correlation model of light beam has rich variety of choices, it cannot be chosen at will because of the non-negative definiteness constraints [43]. To ensure that the constructed correlation structure is reasonable and physically realizable, the elements of CSD matrix have to be non-negative definite kernels. Let's recall the quadratic form

$$Q=\sum_\alpha\sum_\beta\iint f_\alpha^*(\mathbf{r}_1)f_\beta(\mathbf{r}_2)\Gamma_{\alpha\beta}(\mathbf{r}_1,\mathbf{r}_2)\mathrm{d}\mathbf{r}_1^2\mathrm{d}\mathbf{r}_2^2, \quad (\alpha,\beta=x,y), \tag{2}$$

where $f_\alpha$ is an arbitrary (well-behaving) function. Then, the non-negative constraint means that, for any choice of $f_x$ and $f_y$, $Q$ must be non-negative. However, to make sure that $Q$ meets the conditions for a given form of $\Gamma_{\alpha\beta}$ is quite difficult, because of the arbitrariness in the choice of $f_x$ and $f_y$. A sufficiency condition for satisfying the non-negative constraint in Eq. (2) has recently been proposed [43]. It is shown that a well-defined form for the elements of CSD with the help of the reproducing kernel Hilbert spaces can be expressed as an integral

$$\Gamma_{\alpha\beta}(\mathbf{r}_1,\mathbf{r}_2)=\iint p_{\alpha\beta}(\mathbf{v})K_\alpha^*(\mathbf{r}_1,\mathbf{v})K_\beta(\mathbf{r}_2,\mathbf{v})\mathrm{d}\mathbf{v}^2, \quad (\alpha,\beta=x,y), \tag{3}$$

where $\mathbf{v}\equiv(v_x,v_y)$, $p_{\alpha\beta}$ is an non-negative weighting functions and can consist of a set of delta functions with positive coefficients. $K_\alpha$ is an arbitrary kernel and can be regard as the physically available response functions associated with the electric field components $K_\alpha$. On inserting from Eq. (3) into Eq. (2), for any non-negative definiteness constraint of $Q$, $p_{\alpha\beta}$ should satisfy the following inequality [44]

$$p_{xx}(\mathbf{v})\gg 0,\ p_{yy}(\mathbf{v})\gg 0,\ \text{and}\ p_{xx}(\mathbf{v})p_{yy}(\mathbf{v})-|p_{xy}(\mathbf{v})|^2\gg 0, \tag{4}$$

for any $\mathbf{v}$. As a consequence, the elements of CSD matrix constructed by means of the recipe given by Eq. (3) are necessarily non-negative definite, namely, it is physically realizable, as far as Eq. (4) is satisfied.

Since $K_\alpha$ represents arbitrary kernels of linear transformation associated with $\mathbf{v}$, it has various choices such as Fourier, Laplace, Hankel, mellin, and etc.,. Here, for simplicity, let us suppose kernel $K_\alpha$ to be a conventional Fourier-like form,

$$K_\alpha(\mathbf{r},\mathbf{v}) = U_\alpha(\mathbf{r})\exp[-2\pi i g_\alpha(\mathbf{r})\cdot\mathbf{v}], \tag{5}$$

where $U_\alpha(\mathbf{r})$ is possible complex profile function and $g_\alpha$ is arbitrary vectorial real function. On inserting Eq. (5) into Eq. (3), we get the following form

$$\Gamma_{\alpha\beta}(\mathbf{r}_1,\mathbf{r}_2) = U_\alpha^*(\mathbf{r}_1)U_\beta(\mathbf{r}_2)\tilde{p}_{\alpha\beta}\left[g_\beta(\mathbf{r}_2) - g_\alpha(\mathbf{r}_1)\right]. \tag{6}$$

where the tilde represents the Fourier transform. It is seen that $\tilde{p}_{\alpha\beta}$ coincides with the source complex degree of coherence (DOC). Since $g_\alpha$ can be chosen as will, the elements $\Gamma_{\alpha\beta}$ can have rather sophisticated structures. Let us consider the simplest case, for instance $g_\alpha(\mathbf{r}) = \mathbf{r}$. In comparison with Eq. (3) and Eq. (6), one finds that the source correlation function is the Fourier transform of weighting function $p_{\alpha\beta}$.

It is worthwhile to recall a typical case that the correlation structure has a Gaussian form (known as Schell-model (SM)) [33–45]. Then, the weighting function $p_{\alpha\beta}$ has the following form

$$p_{\alpha\beta}^{(SM)}(\mathbf{v}) = 2\pi\sigma_0^2 \exp[-2\pi^2\sigma_0^2\mathbf{v}^2]. \tag{7}$$

In the following, let us further construct $p_{\alpha\beta}$ to have a $M\times N$ radial array structure with the help of a periodic comb function,

$$p_{\alpha\beta}^{(O)}(v_x,v_y) = \frac{2\pi\sigma_0^2}{C_o}\sum_{n=0}^{N}\sum_{m=1}^{M}\exp[-2\pi^2\sigma_0^2(v_x + mr_0\cos\varphi_n)^2]\exp[-2\pi^2\sigma_0^2(v_y + mr_0\sin\varphi_n)^2], \tag{8}$$

where $C_o = M(2N+1)$ and $\varphi_n = 2\pi n/(N+1)$, $r_0$ is the radial of the inner ring. On the other hand, since profile function can be chosen at will, let us take into account a typical RP form [12, 13],

$$U_\alpha^*(\mathbf{r}_1)U_\beta(\mathbf{r}_2) = \frac{\alpha_1'\beta_2'}{w_0^2}\exp\left[-\frac{\mathbf{r}_1^2 + \mathbf{r}_2^2}{4w_0^2}\right],\ (\alpha,\beta=x,y), \tag{9}$$

where $w_0$ denotes the transverse beam size. On inserting Eqs. (8) and (9) into Eq. (6), we obtain the following expressions for the elements of CSD matrix

$$\Gamma^{(O)}_{\alpha\beta}(\mathbf{r}_1,\mathbf{r}_2) = \sum_{n=0}^{N}\sum_{m=1}^{M}\frac{\alpha'_1\beta'_2}{NMw_0^2}\exp\left[-\frac{\mathbf{r}_1^2+\mathbf{r}_2^2}{4w_0^2}-\frac{(\mathbf{r}_2-\mathbf{r}_1)^2}{2\sigma_0^2}\right]\exp\left\{i2\pi mr_0\left[\cos\varphi_n(x'_2-x'_1)+\sin\varphi_n(y'_2-y'_1)\right]\right\}. \quad (10)$$

One can notice that Eq. (10) features a RP beam with a Schell-model array correlated (SMAC) function of radial symmetry. By evaluating Eq. (8), it is easy to find

$$p_{xx}(\mathbf{v}) = p_{yy}(\mathbf{v}) = p_{xy}(\mathbf{v}) > 0, \text{ and } |p_{xy}(\mathbf{v})|^2 \equiv p_{xx}(\mathbf{v})p_{yy}(\mathbf{v}). \quad (11)$$

According to the Eq. (10), it is clear that Eq. (4) is satisfied for any $\mathbf{v}$. Furthermore, for this new type of non-uniformly correlated RP beam, the beam condition for radiation generated by such source can refer to Refs. [49–52], and it is derived as

$$\frac{1}{4w_0^2}+\frac{1}{\sigma_0^2} \ll \frac{2\pi^2}{\lambda^2}(1-2\lambda Mr_0)^2. \quad (12)$$

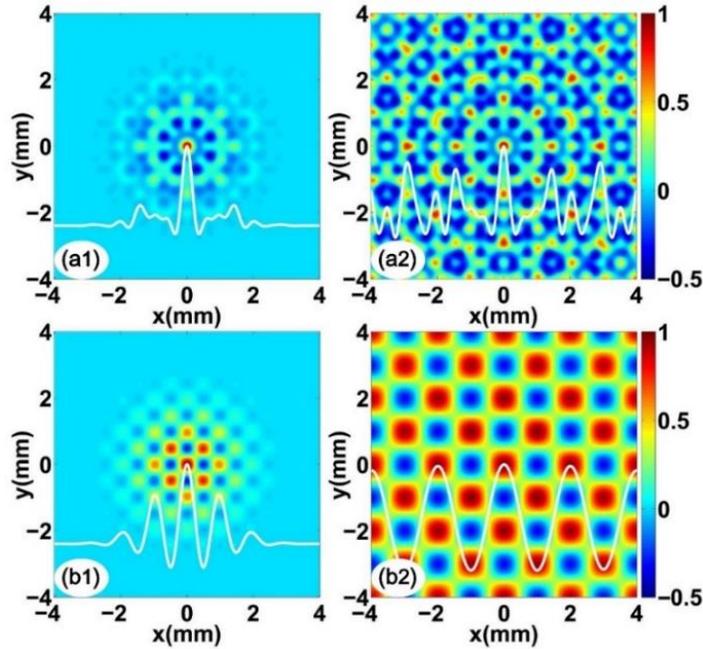

FIG. 1. Contour graphs of the source spectral DOCs and the corresponding cross lines for radial and rectangular SMAC RP beams as a function of $\sigma_0$. (a1)-(b1): $\sigma_0 = 1\text{mm}, 2a = r_0 = 1\text{mm}$, (a2)-(b2): $\sigma_0 = 10\text{mm}, a = 2r_0 = 1\text{mm}$.

Figure 1 shows the source spectral DOCs and the corresponding cross lines for radial and rectangular SMAC RP beams as a function of $\sigma_0$. The parameters are chosen to be $\lambda = 632.8\text{nm}, w_0 = 1\text{mm}, M = 2, N = 7, M' = N' = 1$, and $\sigma_0$ represents the initial coherence width. When $M = N = M' = N' = 0$, the initial beams will reduce to a conventional RP SM beam. One can see in Fig. 1 that the array structures are determined by the comb correlation function, and the region of array is controlled by the Gaussian correlation functions. As the increase of the value $\sigma_0$, the effective correlation array region gradually expand. Due to the nonconventional correlation structure, light beams thus exhibit extraordinary modulation features compared to those beams generated from SM sources during propagation. We shall emphasize that the propagation-induced statistics changes are completely due to the correlation structure and do not result from any deterministic phase modulation.

In the following, let us consider the paraxial propagation of this beam through a stigmatic ABCD optical system. The components of spectral intensity can be evaluated by the generalized Collins formula [13, 33]

$$\Gamma_{\alpha\beta}^{(O)}(\boldsymbol{\rho},\boldsymbol{\rho}) = \frac{1}{\lambda^2 B^2} \iint \Gamma_{\alpha\beta}^{(O)}(\mathbf{r}_1,\mathbf{r}_2) \exp\left[-\frac{ikA}{2B}(\mathbf{r}_1^2 - \mathbf{r}_2^2)\right] \exp\left[\frac{ik}{B}(\mathbf{r}_1 \cdot \boldsymbol{\rho} - \mathbf{r}_2 \cdot \boldsymbol{\rho})\right] d^2\mathbf{r}_1 d^2\mathbf{r}_2, \quad (13)$$

where $\boldsymbol{\rho} \equiv (x, y)$ is the position vector in the receiver plane and $k = 2\pi/\lambda$ denotes the wave number with $\lambda$ being the wavelength. A, B, C, and D are elements of the transfer matrix of the optical system. On substituting from Eq. (10) into Eq. (13), we obtain the following expressions for the components of the spectral intensity in the receive plane

$$\Gamma_{\alpha\alpha}^{(O)}(\boldsymbol{\rho},\boldsymbol{\rho}) = \frac{w_0^{-2}}{C_o \Omega^2} \sum_{n=0}^{N} \sum_{m=1}^{M} \left[\frac{B^2}{k^2 \sigma_0^2} + \left(1 - \frac{w_0^{-2} B^2}{k^2 \sigma_0^2 \Omega}\right) \xi_\alpha^2\right] \exp\left(-\frac{\xi^2}{2 w_0^2 \Omega}\right), \quad (\alpha = x, y), \quad (14)$$

$$\Gamma_{xy}^{(O)}(\boldsymbol{\rho},\boldsymbol{\rho}) = \Gamma_{yx}^{(O)*}(\boldsymbol{\rho},\boldsymbol{\rho}) = \frac{w_0^{-2}}{C_o \Omega^2} \sum_{n=0}^{N} \sum_{m=1}^{M} \xi_x \xi_y \exp\left(\frac{-\xi^2}{2 w_0^2 \Omega}\right) \left(1 + \frac{B^2}{k^2 \sigma_0^4 \Omega} - \frac{w_0^{-2} B^2}{2 k^2 \sigma_0^2 \Omega}\right), \quad (15)$$

where

$$\xi_\alpha = \alpha - \lambda m r_0 B \cos\varphi_n, \quad \alpha = (x, y), \quad \text{and} \quad \Omega = A^2 + \frac{B^2}{4 k^2 w_0^4}\left(1 + \frac{4 w_0^2}{\sigma_0^2}\right). \quad (16)$$

By applying a coordinate transform, we can further obtain the following components of spectral intensity for a $(2M'+1) \times (2N'+1)$ rectangular SMAC RP beam,

$$\Gamma_{\alpha\alpha}^{(\perp)}(\boldsymbol{\rho},\boldsymbol{\rho}) = \frac{w_0^{-2}}{C_\perp \Omega^2} \sum_{m=-M'}^{M'} \sum_{n=-N'}^{N'} \left[\frac{B^2}{k^2 \sigma_0^2} + \left(1 - \frac{w_0^{-2} B^2}{k^2 \sigma_0^2 \Omega}\right) \zeta_\alpha^2\right] \exp\left[-\frac{w_0^{-2}}{2\Omega}(\zeta_x^2 + \zeta_y^2)\right], \quad (\alpha = x, y), \quad (17)$$

$$\Gamma_{xy}^{(\perp)}(\boldsymbol{\rho},\boldsymbol{\rho}) = \Gamma_{xy}^{(\perp)*}(\boldsymbol{\rho},\boldsymbol{\rho}) = \frac{w_0^{-2}}{C_\perp \Omega^2} \sum_{m=-M'}^{M'} \sum_{n=-N'}^{N'} \zeta_x \zeta_y \left(1 + \frac{B^2}{k^2 \sigma_0^4 \Omega} - \frac{w_0^{-2} B^2}{2 k^2 \sigma_0^2 \Omega}\right) \exp\left[\frac{-w_0^{-2}}{2\Omega}(\zeta_x^2 + \zeta_y^2)\right], \quad (18)$$

with

$$\zeta_x = x - \lambda mB/2a, \quad \zeta_y = y - \lambda nB/2a, \quad \text{and} \quad C_\perp = (2M'+1)(2N'+1). \tag{19}$$

where $a$ is the separation distance. It is easy to find that the realizability condition for a rectangular SMAC RP source is coincide with the radial SMAC RP source, and the beam condition is obtained as follows

$$\frac{1}{4w_0^2} + \frac{1}{\sigma_0^2} \ll \frac{2\pi^2}{\lambda^2}\left[1 - \frac{\lambda}{2a}\cdot \max(M,N)\right]^2. \tag{20}$$

Let us now take into account the focusing vector properties. Here, we assume that the source beam propagates a distance of $l = 150$mm, and then focused by a thin lens with focal length $f = 150$mm. The elements of the transfer matrix between the source plane and the focal plane are given by

$$A = 1 - \frac{z}{f}, \quad B = f, \quad C = -\frac{1}{f}, \quad D = 0. \tag{21}$$

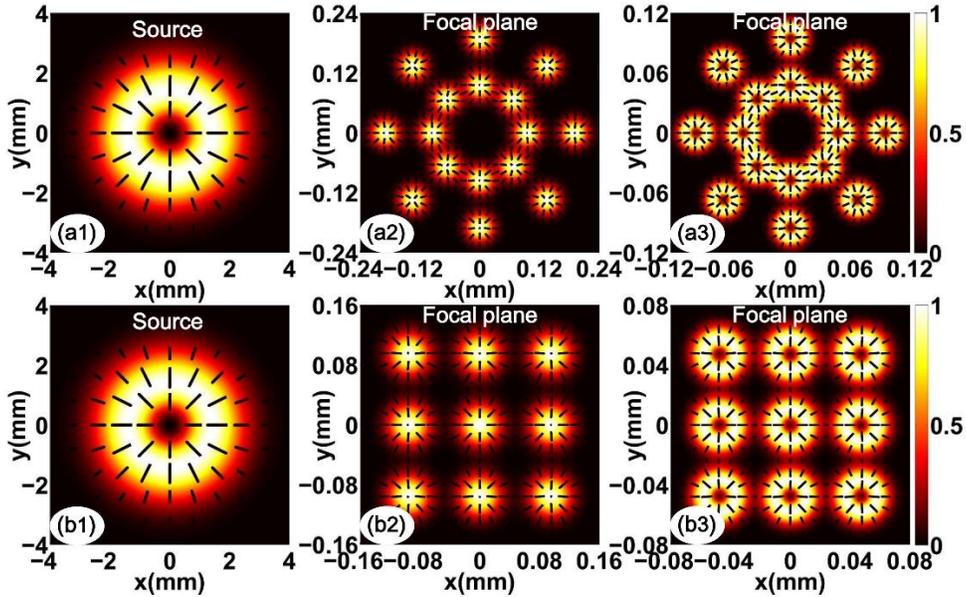

FIG. 2. Contour plots of the SOPs in the source plane for radial and rectangular SMAC RP beams as well as that in the focal plane as a function of $\sigma_0$. (a2)-(b2): $\sigma_0 = 1$mm, $2a = r_0 = 1$mm, (a3)-(b3): $\sigma_0 = 10$mm, $a = 2r_0 = 1$mm.

Figure 2 shows the contour plots of the spectral SOP for radial and rectangular SMAC RP beams in the source plane as well as that in the focal plane. It is clearly seen that the spectral density in the focal plane exhibits array distribution with radial or rectangular symmetry. This means that the Fourier-like relation between the initial correlation function and the far-field spectral intensity distribution is still valid for vector beams [56]. At the same time, each beamlet evolves from a Gaussian shape into a dark-hollow profile when the value of $\sigma_0$ increases. More importantly, the initial overall polarization structure is completely broken causing by the nonconventional source correlation structure, and each beamlet reproduces the initial SOP and displays RP structure. This implies the polarization modulation and amplitude modulation are not independent features that can be simultaneously manipulated by means of the correlation structure modulation. It turns out that this reproduction resulted from the source correlation structure is also applicable to other vector beams such as generic CV beams. However, the present example for modulating a vector beam is more than just a simple split (as it might appear) of initial SOP, and can be more complicated.

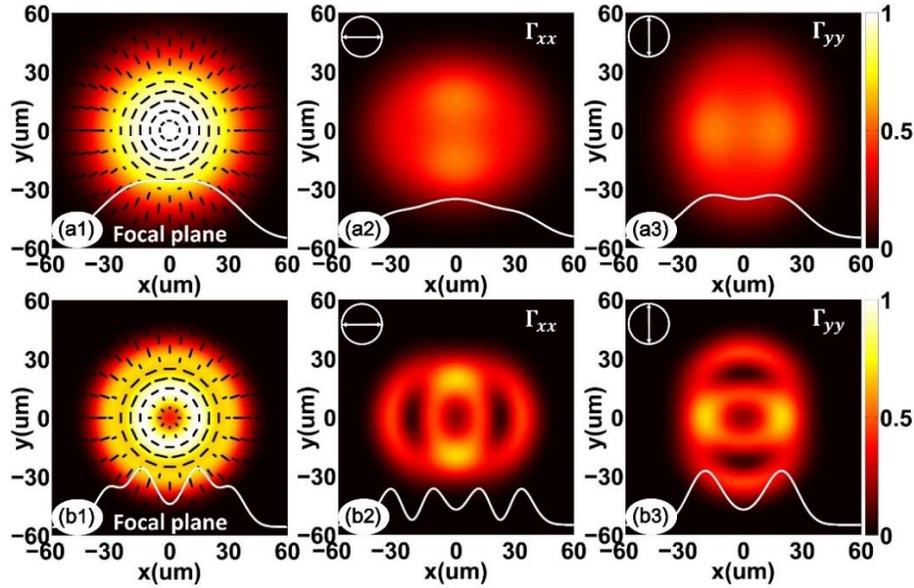

FIG. 3. Contour plots of the SOP, corresponding intensity components and the cross lines of a radial SMAC RP beam in the focal plane with different $\sigma_0$. (a1)-(a3): $\sigma_0 = 1$mm, (b1)-(b3): $\sigma_0 = 10$mm.

Figure 3 illustrates the SOP and the corresponding intensity components of a radial SMAC RP beam in the focal plane as a function of $\sigma_0$. Previous studies show that a separable weighting function $p_{\alpha\beta}$ in the Cartesian coordinate system leads to a rectangle-like far-field intensity profile [45, 56]. However, our results show that this is not a sufficient condition and a circular beam can be obtained as well, although $p_{\alpha\beta}$ is given under the Cartesian coordinate system. Different from the polarization reproduction presented in Fig. 2, it can be seen that the SOP in Fig. 3 displays a more complicated structure. In the case of lower value of coherence width such as $\sigma_0 = 1$mm, one finds from Figs. 3(a1)–3(a3) that the most region of the beam is azimuthally polarized, except the edge portion displays a RP structure. However, as the increase of the coherence width such as $\sigma_0 = 10\text{mm}$, Fig. 3(b1) shows that the SOP in the central region evolves into RP structure. Furthermore, the corresponding intensity components, as presented in Figs. 3(a2), 3(a3), 3(b2) and 3(b3) clearly demonstrate this variation in polarization state. As a remark, further investigations on the correlation-induced polarization state change are helpful for exploring new vector beams with unexpected features.

To demonstrate more about this method, Fig. 4 illustrates an OK-shaped RPA beam in the focal plane based on the special designed initial correlation structures. Once again, we find that not only the array structure can be arbitrarily designed, but also the SOP can be precisely controlled by means of the initial correlation structure. The ability of generation arbitrary vector array beam can find widespread applications in particle manipulation and optical engineering. By using a similar approach, Fig. 5 demonstrates an optical cage intensity profile around the focal region. In particular, the size of optical cage can be effectively regulated by taking advantage of a thin lens. Such an optical cage is essential needed for non-destructive manipulation of biological cells and capture of metal nanoparticles [2, 3]. Finally, it is important to point out that the correlation structures mentioned above can be generated experimentally by means of a liquid crystal spatial light modulator or a method proposed recently in Refs. 44 and 45.

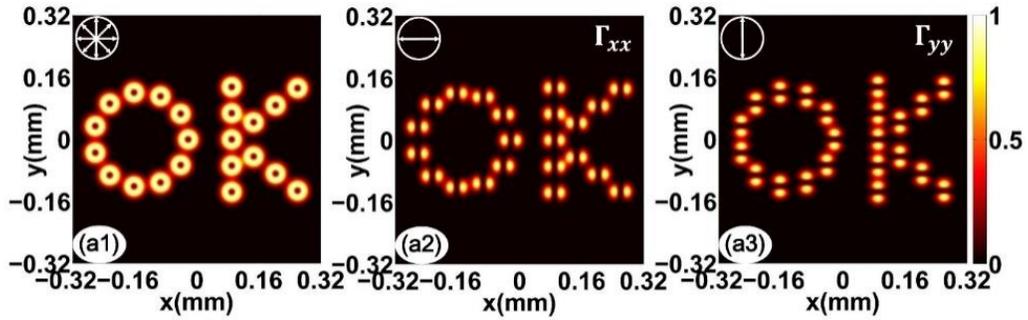

FIG. 4. Contour plots of an OK-shaped RPA beam and its corresponding intensity components in the focal plane.

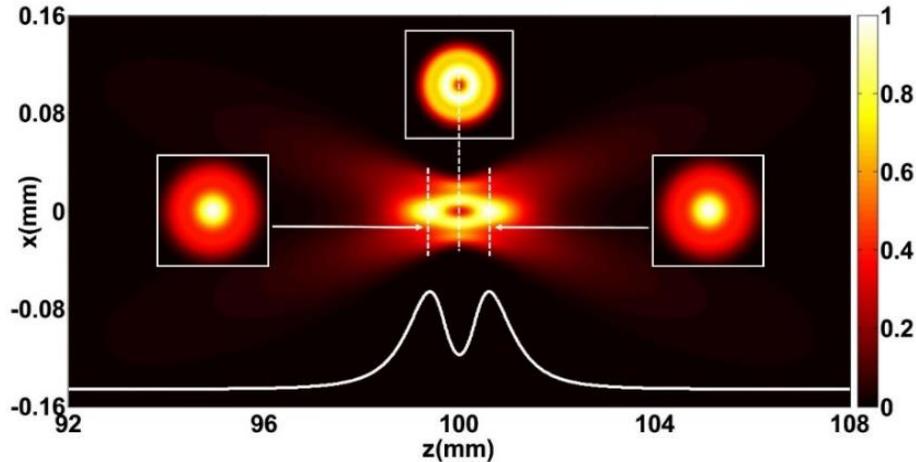

FIG. 5. An optical cage is derived around the focal region.

In conclusion, we have demonstrated a convenient approach for simultaneously manipulating the amplitude and polarization of light beams by means of the design of the correlation structure. As an illustration, by constructing a periodic correlation structure, a RP beam is focused to an arbitrary RPA in the focal plane. The realizability conditions for the correlation structures and the far-field beam condition are established. It has been shown that the far-field spectral intensity distribution and the SOP can be effectively controlled by modulating the initial coherence width. Furthermore, correlation structures give rise to a tunable OK-shaped RPA beam and an optical cage are demonstrated. The ability to appropriately manipulate amplitude and polarization of light beams by designing the correlation structure gives an incentive for the generation of more complex vector beams such as high-order vector beams, generic CV beams or Airy beams and promises important supports in the fields of imaging, laser machining, and optical tweezers.

**ACKNOWLEDGMENTS**

The author's research is supported by the National Natural Science Foundation of China under Grant No. 11504172, the Natural Science Foundation of Jiangsu Province under Grant No. BK20150763, and the project of the Priority Academic Program Development (PAPD) of Jiangsu Higher Education Institutions.